\documentclass[traditabstract]{aa}  
\usepackage{graphicx,amsmath}
\usepackage{txfonts}

\begin{document}
\title{
Hint of 150 MHz radio emission \\ 
from the Neptune-mass extrasolar transiting planet HAT-P-11b
\thanks{Data for these observations can be retrieved electronically 
from the GMRT archive server {\tt http://ncra.tifr.res.in/\symbol{126}gmrtarchive} 
and by request to {\tt archive@gmrt.ncra.tifr.res.in}.}
    }

   \author{
A.~Lecavelier des Etangs\inst{1,2}
 \and
 S.~K.~Sirothia\inst{3}
 \and
 Gopal-Krishna\inst{3}       
 \and
 P.~Zarka\inst{4}
   }
   
\titlerunning{150\,MHz radio emissions from the Neptune-mass planet HAT-P-11b}

\offprints{A.L. (\email{lecaveli@iap.fr})}

   \institute{
   CNRS, UMR 7095, 
   Institut d'Astrophysique de Paris, 
   98$^{\rm bis}$ boulevard Arago, F-75014 Paris, France
   \and
   UPMC Univ. Paris 6, UMR 7095, 
   Institut d'Astrophysique de Paris, 
   98$^{\rm bis}$ boulevard Arago, F-75014 Paris, France
   \and
   National Centre for Radio Astrophysics, TIFR, 
   Post Bag 3, Pune University Campus, Pune 411007, India
   \and
   LESIA, Observatoire de Paris, CNRS, UPMC, Universit\'e Paris Diderot, 
   5 Place Jules Janssen, 92190 Meudon, France
}   
   \date{} 
 
  \abstract
{Since the radio-frequency emission from planets is expected to 
be strongly influenced by their interaction with the magnetic 
field and corona of the host star, the physics of this process 
can be effectively constrained by making sensitive measurements 
of the planetary radio emission. Up to now, however, numerous 
searches for radio emission from extrasolar planets at radio 
wavelengths have only yielded negative results. Here we report 
deep radio observations of the nearby Neptune-mass extrasolar 
transiting planet \object{HAT-P-11b} at 150\,MHz, using the Giant Meterwave Radio 
Telescope (GMRT). On July 16, 2009, we detected a 3$\sigma$ emission
whose light curve is consistent with an eclipse 
when the planet passed behind the star. This emission is 
at a position 14\arcsec\ from 
the transiting exoplanet's coordinates; thus, with a 
synthetized beam of FWHM$\sim$16\arcsec, the position uncertainty 
of this weak radio signal encompasses the location of  \object{HAT-P-11}. 
We estimate a 5\% false positive probability that the observed 
radio light curve mimics the planet's eclipse light curve. 
If the faint signature is indeed a radio eclipse event associated
with the planet,
then its flux would be 3.87 mJy$\pm$1.29 mJy at 150\,MHz. However, 
our equally sensitive repeat observations of the system on 
November 17, 2010 did not detect a significant signal in the
radio light curve near the same position. This lack of confirmation 
leaves us with the possibility of either a variable planetary 
emission, or a chance occurrence of a false positive signal in our first 
observation. Deeper observations are required to confirm 
this hint of 150\,MHz radio emission from HAT-P-11b. 
  
%

}


   \maketitle
%

\section{Introduction}
\label{Introduction}

The importance of the physics behind star-planet interaction, 
in conjunction with theoretical predictions available for radio emission from 
a large number of extrasolar planets (Grie\ss meier et al.\ 2007), 
has provided the impetus for making searches for decameter- and meter-wavelength 
radio emission from a few carefully selected extrasolar planets 
(Bastian et al.\ 2000; Ryabov et al.\ 2004; Winterhalter et al. 2005; 
George \& Stevens 2007; Lazio \& Farrell 2007; Smith et al.\ 2009; 
Lecavelier des Etangs et al.\ 2009, 2011; Lazio et al.\ 2010a, 2010b). 
Using the scaling laws (Zarka et al.\ 2001; Zarka 2007), radio emission from extrasolar 
planets orbiting close to their host stars could reach 10$^3$ to 10$^5$~times 
the radio output from Jupiter and hence be detectable with the  
most sensitive existing radio telescopes. 
Even at large orbital distances ($\ga$1\,AU), 
radio emissions from rapidly rotating planets orbiting stars with high X-UV brightness
could be detectable (Nichols 2011). 
Nonetheless, in the absence of detection, the theoretical estimates made 
for exoplanetary radio cyclotron emission are based 
on a host of unknowns, such as the stellar wind, coronal density, 
and stellar and planetary magnetic fields.

Up to now, various programs to search for radio emission have given negative results.
The telescopes used include the UTR-2 (10~MHz, $\sigma$$\sim$1.6\,Jy), the VLA 
(74~MHz, $\sigma$$\sim$50\,mJy; 325 MHz, $\sigma$$\sim$0.6\,mJy; 1425 MHz, $\sigma$$\sim$15\,$\mu$Jy), 
and the Giant Meterwave Radio Telescope (GMRT; see review in Lazio et al.\ 2009 and Zarka 2011). 
Interferometric observations of high sensitivity 
and high resolution using the GMRT, 
which is a 30-km baseline Y-shaped array consisting of 30 steerable dishes of 
45-metre diameter each providing an effective area around 30\,000\,m$^2$ (Swarup 1990), 
seem very promising for this purpose. 

The planet \object{HAT-P-11b} is 4.3~Earth radii planet (Deming et al.\ 2011) orbiting a bright 
and metal-rich K4 dwarf with a 4.89~day period on an eccentric orbit (e\,$\approx$\,0.2). 
This transiting planet located 38\,pc away from us was discovered by Bakos et al.\ (2010) with 
ground-based photometry using the HATNet survey (Bakos et al.\ 2004) 
and was confirmed by Dittmann et al.\ (2009). With a mass of about 26~Earth mass, 
this planet belongs to the so-called ``hot-Neptune'' category. 
The host star \object{HAT-P-11} is metal rich ([Fe/H]=+0.31$\pm$0.05). 
Its {\it Kepler} photometry has shown that the
stellar surface has active spotted areas at fixed latitudes (Deming et al.\ 2011; 
Sanchis-Ojeda \& Winn 2011). 

The planet's orbital axis is highly misaligned from the star's rotation axis, with 
a sky projected obliquity of  $\lambda \sim 103$\degr\ (Winn et al.\ 2010; Hirano et al.\ 2011). 
Nonetheless, the orbital plane is seen almost edge-on from the Earth and thus
HAT-P-11\,b is eclipsed by the star for about two hours during each orbital cycle. 
The planet's eclipses can be used to discriminate any planetary emission from 
the stellar emission; 
the same technique has been used extensively for measuring 
planetary infrared emission using various space and ground-based telescopes 
(Deming et al.\ 2005, Charbonneau et al.\ 2005, Sing et al.\ 2008).  

\section{Radio observations at 150 MHz and data analysis}
\label{Observations}

We observed the HAT-P-11 field with the GMRT before, during, and after the planetary eclipse, 
from July 16, 2009, 15h30m~UT to July 17, 1h00m UT. 
The data consist of about 10$^9$ visibility measurements along the 435~baselines 
of the GMRT aperture-synthesis array. The integration time was 2 seconds and the 
central frequency was set at 150 MHz with a bandwidth of 16 MHz. 
The short integration time allows us to construct a radio light curve (see below), which 
is also sensitive to any large radio flares. 

Because the July 2009 run provided a $\sim$3$\sigma$ detection 
near the HAT-P-11 position and a radio light curve consistent with an 
eclipse of the planet HAT-P-11b (Sect.~\ref{Results}), we decided to 
carry out a second observation of the same field. 
This second run was scheduled on November 17, 2010, from 6h00m to 16h00m~UT.
Combining the two observation runs, the total data 
acquisition time on the target field is about 20 hours, 
including the total acquisition for the durations of the
two eclipses (about 4 hours).    

The data reduction was done mainly using the {\tt AIPS++} package. 
We used 3C48 as the primary flux density and bandpass calibrator, taking 
a flux density of 64.42~Jy at 148~MHz.
The phase calibrator was the source 2038+513 which was observed 
repeatedly during the runs. 

While calibrating the data, bad data points were flagged at various stages. 
The data for antennas with relatively large errors in antenna-based gain 
solutions were examined and flagged over certain time ranges. Some 
baselines were flagged based on closure errors on the bandpass calibrator. 
Channel and time-based flagging of the data points corrupted by radio frequency 
interference (RFI) was done by applying a median filter with a $6\sigma$ threshold.
Residual errors above $5\sigma$ were also flagged after a few rounds of 
imaging and self-calibration. 
The final image was made after several rounds of phase self-calibration
and one round of amplitude self-calibration where the data were normalized 
by the median gain found for the entire data.
The final image was also corrected for the primary beam shape taken to be a 
Gaussian with FWHM of 173.83\arcmin\ at 150\,MHz.
 
After the final imaging stage, we generated light curves 
of several emission peaks detected around the HAT-P-11 position: 
one light curve of the 
peak source detected in July 2009 within a synthesized beam from HAT-P-11 
(at such a low signal to noise ratio, the accuracy on the position of a detected 
source is approximately the beam size), and the light curves of 
14 emission peaks with similar S/N and detected around HAT-P-11 
in both observations. These last light curves 
can be used for comparison purposes and validation of the statistics.
To obtain these light curves, we calculated model 
visibilities using the sources detected in the entire field-of-view 
excluding a synthesized beam-wide region centered 
on the desired locations ($\alpha_0, \delta_0$); we then subtracted 
this model from the final calibrated visibility data. The residual 
visibility data (RVD) were then phase-centered on $\alpha_0, \delta_0$ 
and averaged for desired time bins to
generate the light curves.

\begin{figure}[htb]
\begin{center}
\hbox{
 \includegraphics[width=\columnwidth
 , viewport=32 215 444 618, clip
 ]
{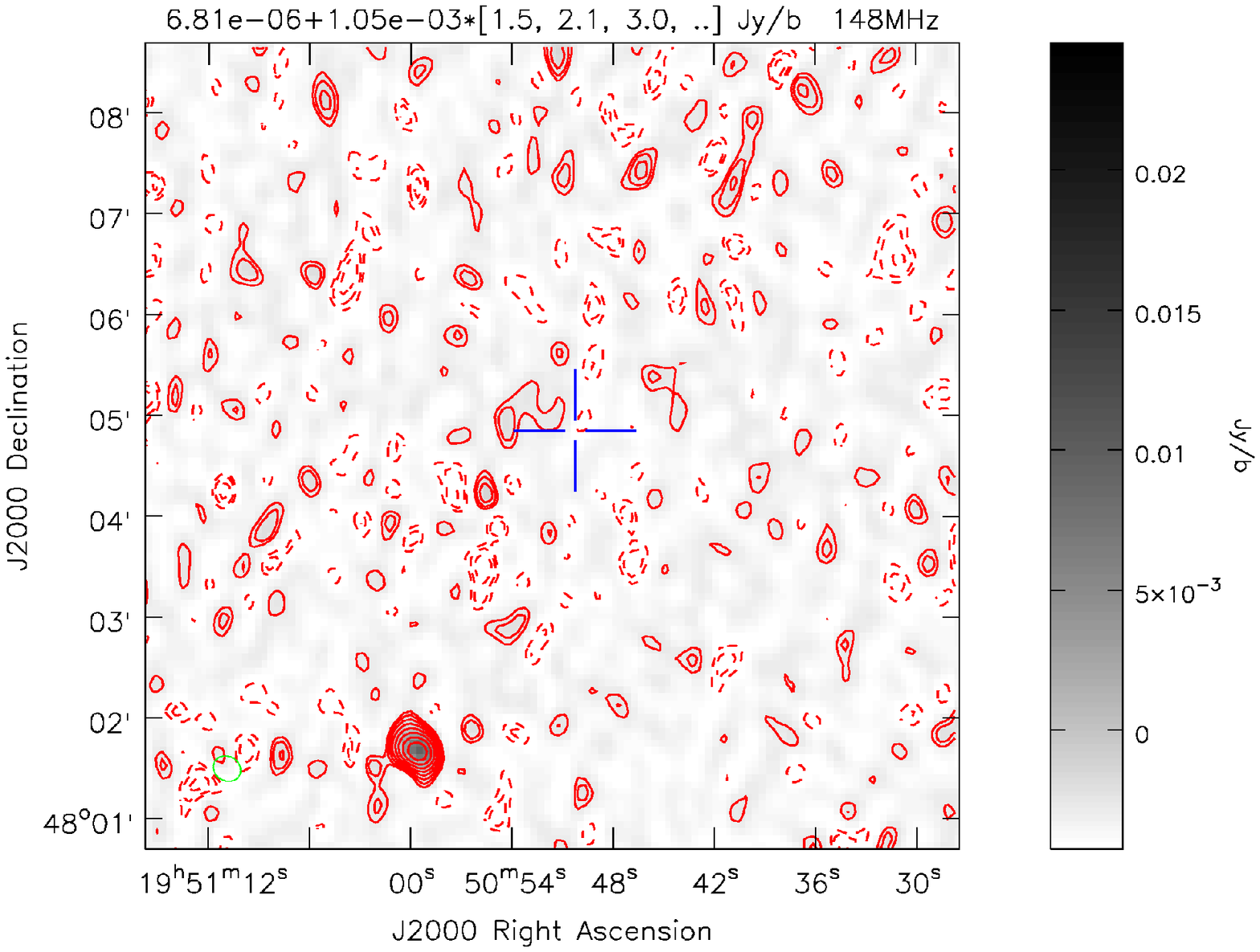}
}
\caption{
GMRT image of the HAT-P-11 field at 150MHz on July 16, 2009. 
The green ellipse in the lower-left corner shows the half-power beam width.
The contour levels given at the top of the image are in units of Jy beam$^{-1}$ and
are defined as mean$+$rms$\times$(n), where values of n are given in the brackets. 
Thus, the first, second, and third contours correspond to 1.5$\times$1.05\,mJy, 
2.1$\times$1.05\,mJy, 3.0$\times$1.05\,mJy, etc.
Negative contours appear as dashed lines.
}
\label{Map 2009 07 16}
\end{center}
\end{figure}

\begin{figure}[htb]
\begin{center}
\hbox{
\includegraphics[angle=90,width=\columnwidth
]{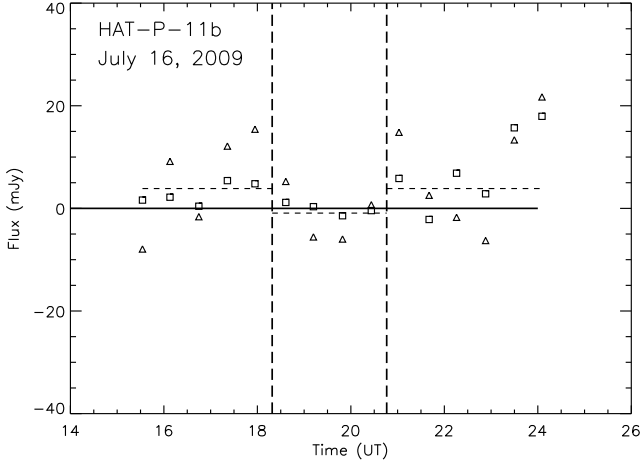}
}
\caption{Time series of the 150\,MHz flux density measured on July 16, 2009, 
in the direction of the radio source near HAT-P-11. The measurements have been rebinned to 
36~minutes.
Triangles and squares correspond to the RR and LL polarizations, respectively.
The two vertical long-dashed lines indicate the beginning and the end of the 
planet's eclipse behind the host star. The dashed horizontal lines show the box-shaped 
eclipse light curve fitted to the data points.}
\label{Light curve 1}
\end{center}
\end{figure}

\begin{figure}[htb]
\begin{center}
\hbox{
\includegraphics[angle=90,width=\columnwidth
]{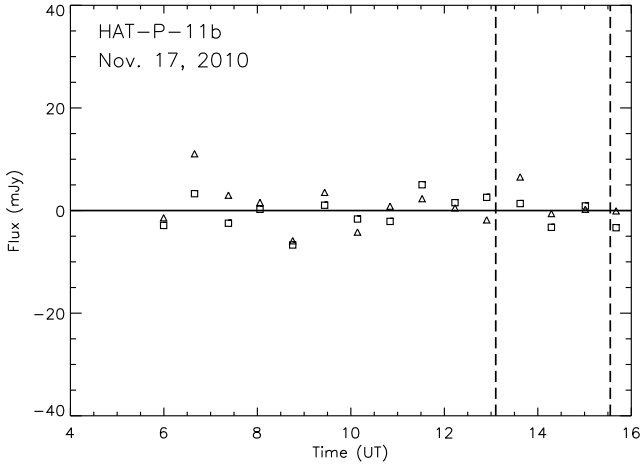}
}
\caption{Same as in Fig.~\ref{Light curve 1}, in the direction of HAT-P-11 on November 17, 2010. 
The mean of the flux density measurements is found to be 0.00$\pm$1.19\,mJy. 
Thus, no emission is detected towards or near HAT-P-11 in these observations.}
\label{Light curve 2}
\end{center}
\end{figure}

\begin{figure}[htb]
\begin{center}
\hbox{
\includegraphics[angle=90,width=\columnwidth
]{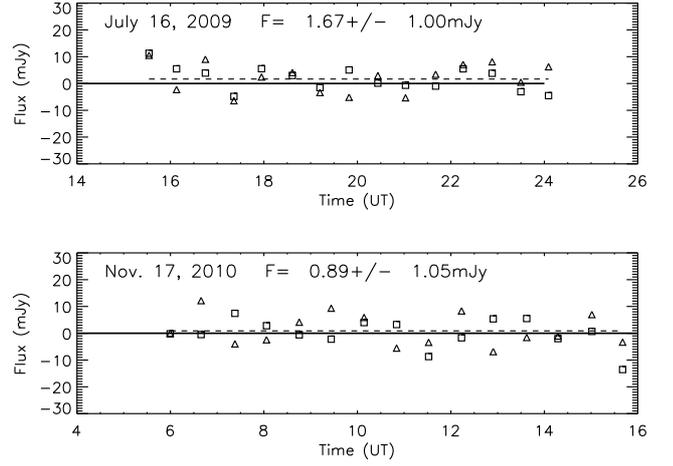}
}
\caption{Time series of the 150\,MHz flux density measured in the direction of one comparison source
at $\alpha$=19h50m32.2s, $\delta$=+48\degr04\arcmin22\arcsec\ (J2000), at 183\arcsec from HAT-P-11. 
The dashed horizontal lines show the mean values of flux density at the two observation epochs.}
\label{Comparison 1}
\end{center}
\end{figure}

\begin{figure}[htb]
\begin{center}
\hbox{
\includegraphics[angle=90,width=\columnwidth
]{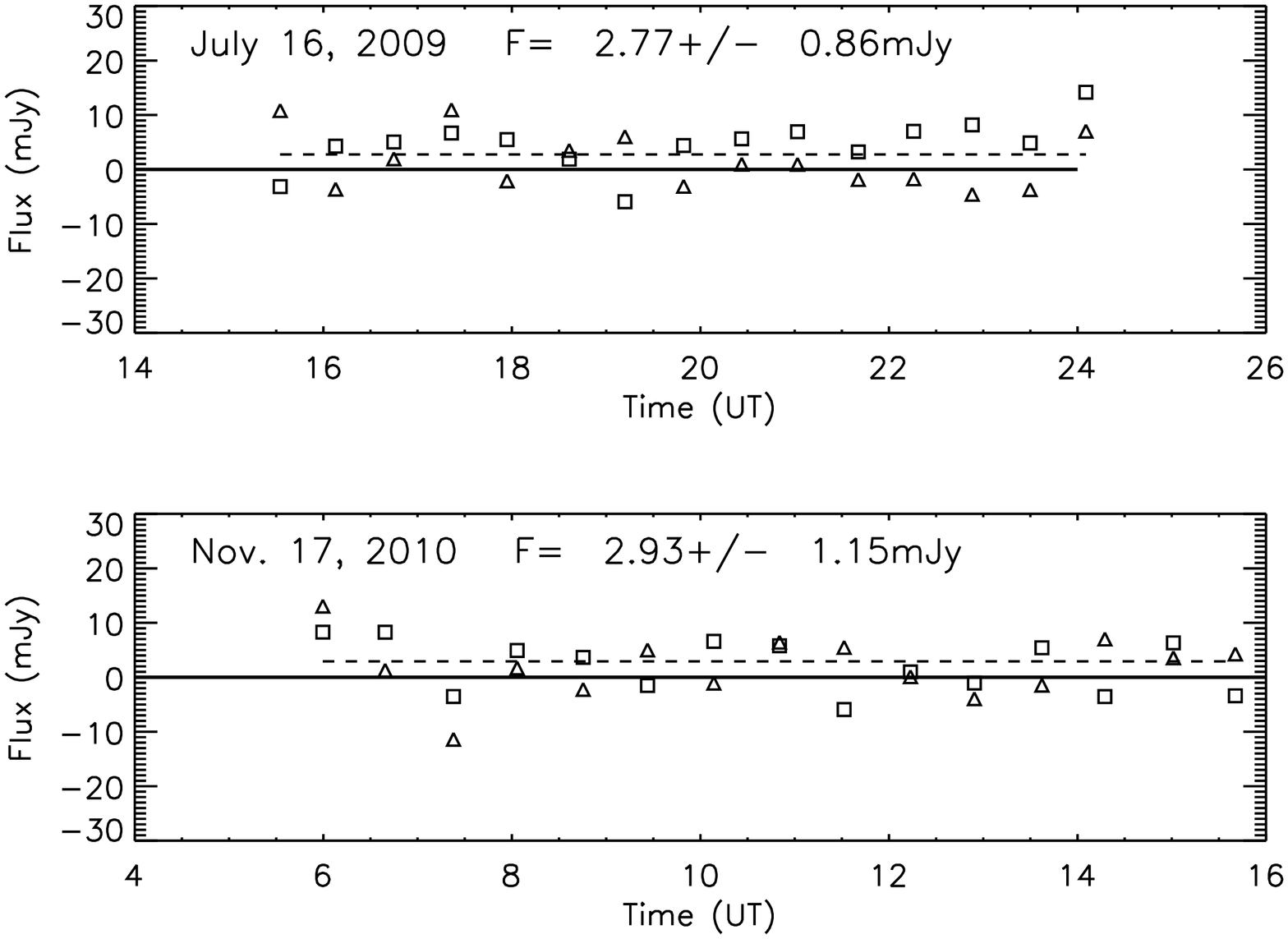}
}
\caption{Same as Fig.~\ref{Comparison 1}, but for the observed peak at $\alpha$=19h50m45.8s, $\delta$=+48\degr04\arcmin10\arcsec\ (J2000), at 61\arcsec from HAT-P-11.}
\label{Comparison 2}
\end{center}
\end{figure}

\begin{figure}[htb]
\begin{center}
\hbox{
\includegraphics[angle=90,width=\columnwidth
]{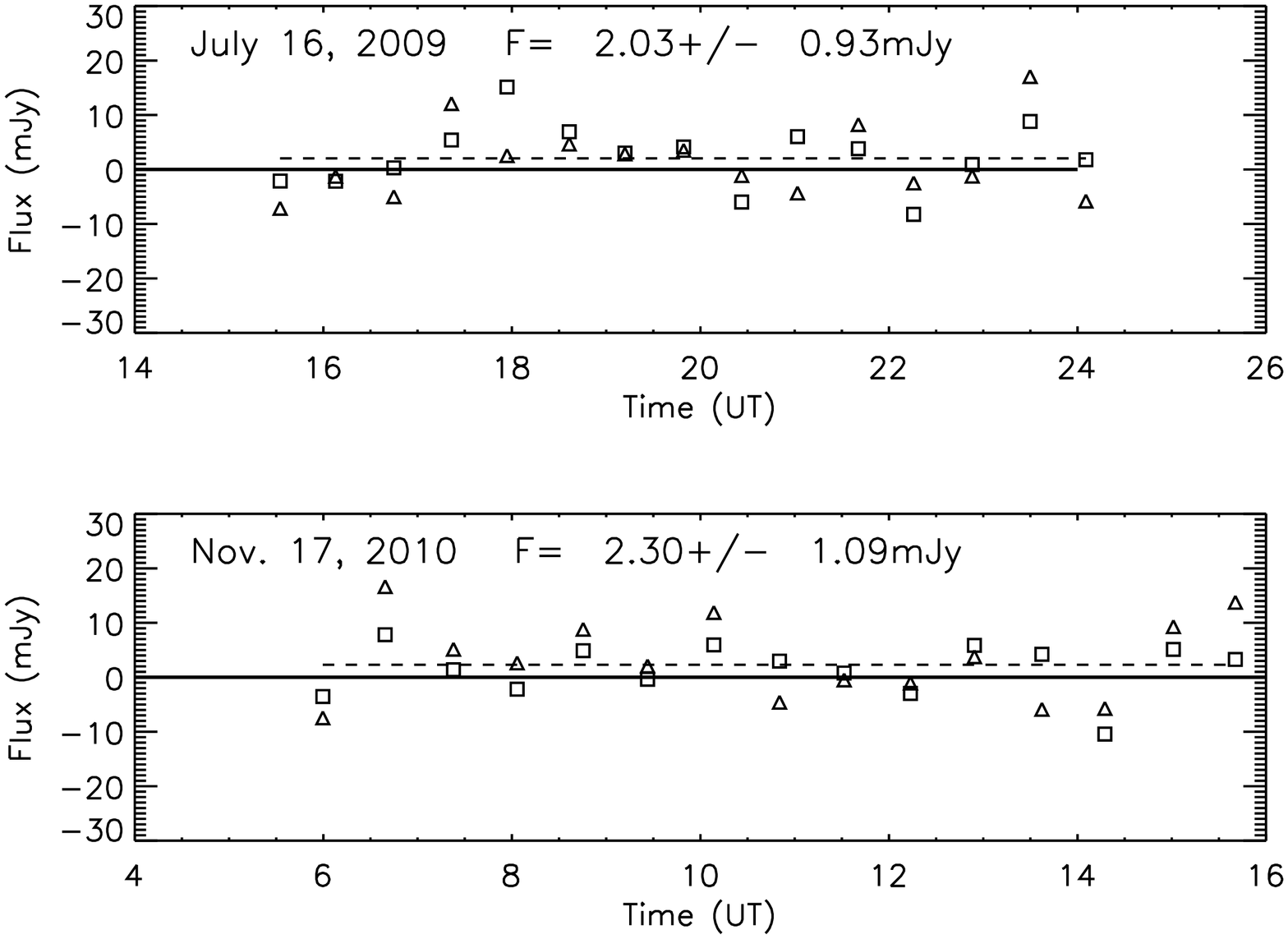}
}
\caption{Same as Fig.~\ref{Comparison 1}, but for $\alpha$=19h50m56.9s, 
$\delta$=+48\degr05\arcmin46\arcsec\ (J2000), at 87\arcsec from HAT-P-11.}
\label{Comparison 3}
\end{center}
\end{figure}

\begin{figure}[htb]
\begin{center}
\hbox{
\includegraphics[angle=90,width=\columnwidth
]{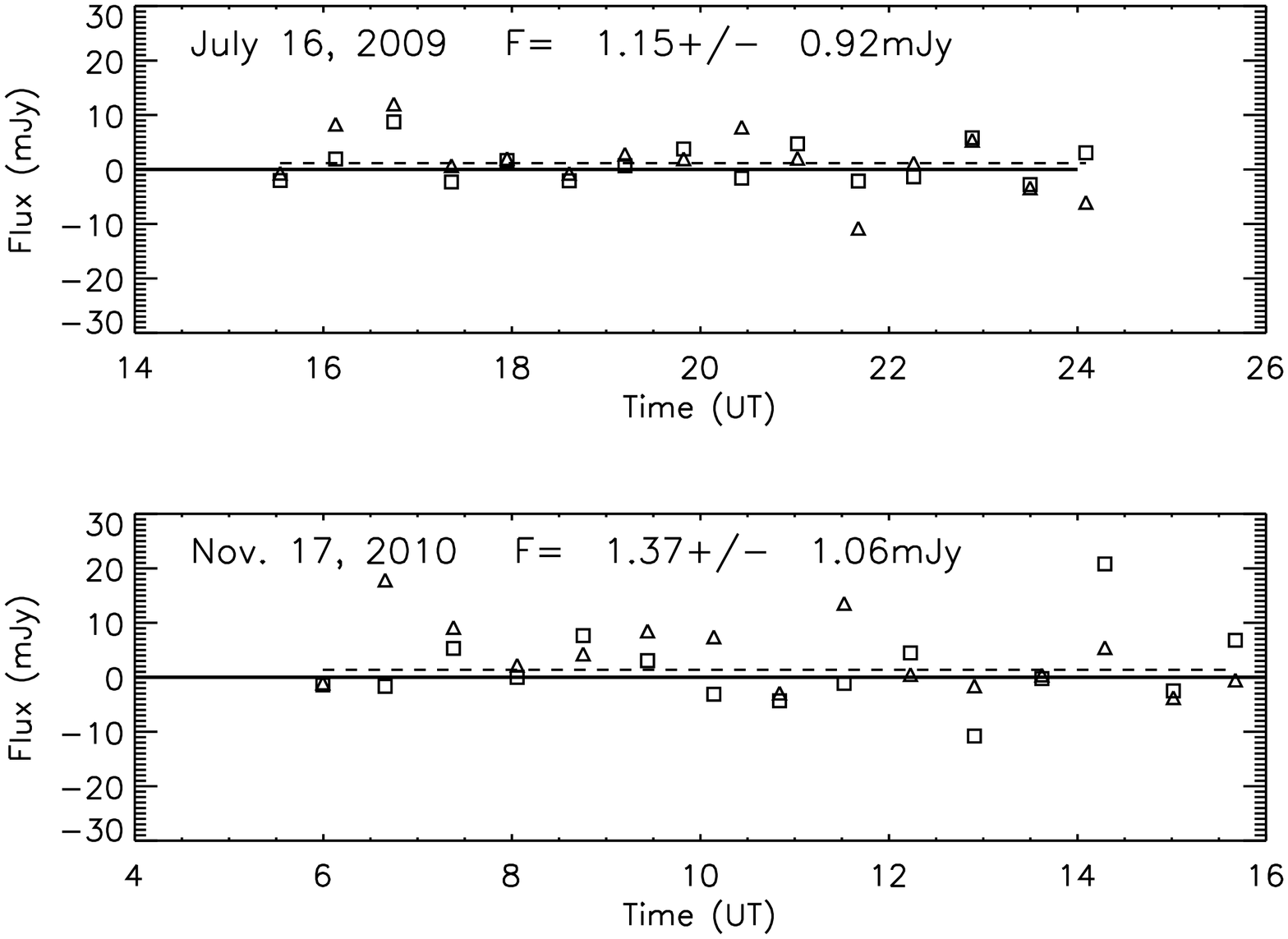}
}
\caption{Same as Fig.~\ref{Comparison 1}, but for $\alpha$=19h50m58.9s, 
$\delta$=+48\degr02\arcmin26\arcsec\ (J2000), at 169\arcsec from HAT-P-11.}
\label{Comparison 4}
\end{center}
\end{figure}

\section{Results}
\label{Results}

\subsection{Observations of July 16, 2009}
\label{Observations of July 16, 2009}

Figure~\ref{Map 2009 07 16} shows the 150\,MHz map of the field around HAT-P-11 obtained
from the observations on July 16, 2009. A faint source is detected at the coordinates 
$\alpha$=19h50m51.0s, 
$\delta$=+48\degr05\arcmin03\arcsec\ (J2000), which are 14\arcsec\ (about a beam width) 
north-east from the HAT-P-11 coordinates 
($\alpha$=19h50m50.247s, $\delta$=+48\degr04\arcmin51.08\arcsec). 
Figure~\ref{Light curve 1} shows the 150\,MHz light curve derived at the source position. 

Over the full integration time, including the two-hour eclipse, 
the source has a mean flux of 2.45$\pm$1.09\,mJy (2.2$\sigma$ detection). 
If we consider only the time outside the eclipse (Fig.~\ref{Light curve 1}), 
a mean flux of 3.87$\pm$1.29\,mJy ($3\sigma$) is measured.
During the eclipse of the planet behind the host star, 
no emission was detected and the mean flux was found to be 
-0.92$\pm$1.94\,mJy, consistent with zero
emission for this time interval. 
The light curve is thus in agreement with the expected light curve 
for a planetary radio emission, eclipsed when 
the planet is known to be behind its host star. 
Thus, a straightforward, albeit heuristic, interpretation of the light curve 
would be a 3$\sigma$ detection of the planet HAT-P-11\,b at 150\,MHz. 

The position of the source exhibiting the eclipse light curve has been measured to be 
14\arcsec\ away from HAT-P-11.
However, for such a faint source the positional accuracy is similar to the beam size of 16\arcsec. 
More precisely, from our experience with GMRT maps at 150~MHz, we derived an empirical 
expression for the positional accuracy on GMRT maps; at 90\% confidence we found 
an accuracy on the position given by $ Beamsize \times (1/SNR + 1/10)$, 
which is in agreement with the same relationship found on the VLA First Survey Catalog 
which reports\footnote{http://sundog.stsci.edu/first/catalogs/readme.html} 
a 90\% confidence positional accuracy of 
$ Beamsize \times (1/SNR + 1/20)$. 
Applying this relationship to our measurement of the emission peak at 2.2$\sigma$ and 
taking into account the 4\arcsec\ pixel size, we find that 
the above measurement of the radio position can be consistent with the HAT-P-11 coordinates. 

To assess the reliability of this detection, we also estimated 
the false-positive detection probability, {\it i.e.}, 
the probability of observing a light curve of this shape 
for a constant radio emission with no real eclipse, 
but only an eclipse mimicked by the random noise along the radio light curve.
This is calculated by estimating the probability that the 
mean flux during the eclipse is lower than zero, as observed. 
For this calculation we used the present estimate 
of the source flux, and the error bars 
for the times outside and during the eclipse. For the case of no real flux variation, 
our analysis shows that the probability of the mean
flux during the eclipse being below zero is only 4.8\%. 
We conclude a 5\% false positive probability of eclipse detection 
in the observed 150\,MHz light curve at the position coincident (within the errors) 
with the planet HAT-P-11b. 

We also computed the $\chi^2$ of two different models fitting the light curve data 
with a 339-second sampling time providing a total of 190 measurements. 
In the first 
single-parameter
model, we assume a constant flux level for the entire light curve, 
corresponding to the hypothesis that the radio source is not the planet and therefore the flux 
density is constant in time. We find a $\chi^2$ of 186.6 for 189 degrees of freedom. 
In the second model, we assume that the source is a planetary emission and the 
flux density during the eclipse is fitted by a second parameter, larger than or equal to zero. 
With this two-parameter model we find a $\chi^2$ of 182.8 
for 188 degrees of freedom. Using the F-test, we conclude that the 
decrease of $\chi^2$ by 3.8 favors the second model with a 95.0\% confidence level, which 
is consistent with the conclusion based on the false positive probability calculated above.

To make the above analysis more rigorous, we also derived the radio light curve for 
14~sources seen within a few arc minutes around HAT-P-11 and detected at both
observation epochs. The measured fluxes of these comparison sources are in the 
range $\sim$1 to 3~mJy. Four examples of light curves are given 
in Figs.~\ref{Comparison 1} to~\ref{Comparison 4}.
We found that the false positive probabilities computed for these directions 
are uniformly distributed between 0 and 1, validating our approach. 
Thus, to a 95\% confidence limit, the detected signal is consistent with
a 150\,MHz radio emission associated with the extrasolar planet HAT-P-11b.

\subsection{Observations of November 17, 2010}
\label{Observations of November 17, 2010}

In the full-synthesis GMRT observations of November 17, 2010,
we did not detect any emission toward HAT-P-11 
or nearby directions within a synthesized beam. 
On this occasion, the nearest detected radio source lies at 27.5\arcsec\ from HAT-P-11, 
and has a flux of 1.78$\pm$1.01\,mJy. The light curve of this source at 150\,MHz 
does not show any eclipse-like signature. 
Therefore, we are led to conclude that 
in these observations of HAT-P-11 we did not detect the radio source and its associated 
eclipse event we had seen on July 16, 2009.

\subsection{False positive or variability?}
\label{False positive or variability ?}

There are two plausible explanations for the discrepancy between the results 
of the two epochs. First, a false positive detection at the first epoch
cannot be completely ruled out given the estimated false positive 
probability of 5\% (Sect.~\ref{Observations of July 16, 2009}).
Moreover, the source is close to the detection limit. So a false detection 
due to noise cannot be excluded at present.

The second interpretation invokes the general variability 
of planetary radio emission at low frequencies.
The variable nature of the radio emission is indeed predicted by models 
including the observed variability of the stellar magnetic field.
For instance, in the case of \object{HD\,189733b} 
Fares et al.\ (2010) showed that the observed variations in the position of the 
so-called ``source surface'' (the stellar surface beyond 
which the magnetic field becomes purely radial)
produce variations in the expected radio flux. 
Thus, in the case of HAT-P-11, if the 
discrepancy between the results of July 2009 and November 2010 is due
to variability of the radio emission, we may have actually detected sporadic emission 
from the extrasolar planet at 150\,MHz on July 16, 2009.
However, the variable nature of the emission precludes a claim of detection; 
we are left with a hint, and we must wait for further, 
possibly even more sensitive observations. 

\section{Ephemeris uncertainties}

The characteristics and ephemeris of the planet's orbit are needed to 
compute the eclipse light curve. Here we used the orbit's eccentricity 
$e$ and the longitude of periastron $\omega$ provided in the discovery 
paper by Bakos et al.\ (2010): $e\cos \omega$=0.201$\pm$0.049 and 
$e\sin \omega$=0.051$\pm$0.092. The more recent observations of transits 
of HAT-P-11\,b using the Kepler satellite provide new ephemeris 
for the transit central time $T_0$ and orbital period $P$ (Deming et al.\ 2011, 
Sanchis-Ojeda\& Winn 2011). 
We used the following values given in Deming et al.\ (2011): $T_0$=2454605.89155 (BJD) 
and $P$=4.8878018~days. The difference between the eclipse time calculated using these 
new values and the time calculated using the $T_0$ and $P$ given in Bakos et al.\ is 
not more than 1.5~minutes. 

More importantly, the new measurements using the Kepler photometric data yield a smaller 
impact parameter $b_1=a\cos i / R_p$, where $a$ is the semi-major axis, $i$ the orbit's 
inclination, and $R_p$ the planet's radius. A smaller impact parameter leads to a longer 
eclipse duration. Using the new values, we found an eclipse duration of 2~hours and 
27~minutes, which is 38 minutes longer than the previous estimates using the
Bakos et al.\ (2010) impact parameter. 

By far, the largest uncertainty in computing the eclipse time arises from the uncertainty 
in the orbital eccentricity and longitude of the periastron, the uncertainty 
reaching up to $\pm$3h30m in the extreme cases. Unfortunately, the eclipse amplitude 
in the infrared may be too faint to be detected using the Spitzer Space Telescope
(Deming et al.\ 2011). Only new radial velocity measurements in the coming years 
may improve the accuracy of the orbital eccentricity. Here, we used the best published 
values given by Bakos et al.\ (2010) and used the eclipse phase $\phi$=0.628, 
where $\phi$=0 corresponds to the central transit of the planet in front of the host star. 

\section{Discussion}

\subsection{Radio emission frequency}

The principal mechanism put forward for planetary radio emission is 
electron-cyclotron maser radiation. 
It occurs at the local gyrofrequency $f_g=2.8\left({B_p}/{\rm 1\,G}\right) {\rm MHz}$,
where $B_p$ is the planet's magnetic field. 
The detection at a frequency of 150\,MHz would then correspond to 
a planetary magnetic field of around 50~G. 

We recall that the Jovian magnetic field estimated from the radio frequency 
cut-off in its cyclotron emission is around 14~G. 
Thus, the present tentative radio detection 
requires a magnetic field for HAT-P-11\,b ($\sim$50~G) 
that is of the same order of magnitude, albeit slightly stronger than that of Jupiter. 
We have no clues to the possible origin of a magnetic field of this strength, 
although we can emphasize here the particularities of this planet which has 
a significantly higher density than Jupiter, and orbits a metal-rich 
star ([Fe/H]=+0.31$\pm$0.05). The planet HAT-P-11\,b is consequently expected to have 
a high abundance of heavy elements and a large core mass (Guillot et al.\ 2006).
Taken all together, its peculiarities (eccentric and highly inclined orbit, 
metal-rich host star, short orbital period, etc.) make this planet 
different enough from any planet in our own solar system that it is difficult 
to make a reasonable guess of its magnetic field strength. 
Hence the requirement of a magnetic field 3 to 4~times Jupiter's seems 
quite plausible. 

\subsection{Emission flux}

Several mechanisms for exoplanet radio emissions have been considered by 
Grie\ss meier et al.\ (2007): stellar wind (SW)-magnetosphere magnetic interaction, 
SW-magnetosphere kinetic interaction, 
and coronal mass ejection-magnetosphere kinetic interaction. 
Application to HAT-P-11\,b of the scaling law extrapolations of Grie\ss meier et al.\ (2007)
lead to predicted radio flux densities of 14~mJy for the SW-magnetosphere magnetic interaction, 
1~mJy for the coronal mass ejection (CME)-magnetosphere kinetic interaction, and $<<$1~mJy for the 
SW-magnetosphere kinetic interaction (note that these are mere orders of magnitude). 
The first two values are of the same order of magnitude 
as the $\sim$4~mJy flux density tentatively detected here. 

Alternatively, if the stellar magnetic field is strong enough, 
the planet may induce electron precipitations and thus radio emission 
in the star's magnetic field (Zarka et al.\ 2001; Zarka 2007). 
This may be the case for HAT-P-11\,b whose semi-major
axis is only 0.053\,au. The stellar magnetosphere 
could even extend beyond the planet's orbit.
In the framework of the Jardine \& Cameron (2008) model
in which the radio emission is caused by interaction of the planetary magnetosphere 
with the stellar corona in which it is embedded, we can derive typical values 
for the resulting radio signal from HAT-P-11\,b.  
Because HAT-P-11 is a slowly rotating star (Bakos et al.\ 2010) with a rotation period 
of 30.5$^{+4.1}_{3.2}$\,days (Sanchis-Ojed \& Winn 2011),  
the planet's velocity relative to the star's magnetosphere is $v_p - v_*$=99\,km\,s$^{-1}$
({\it i.e.}, three times higher than that of HD\,189733b). 
Taking a 10\% efficiency for conversion of the power of the accelerated 
electrons into radio emission within an emission beam solid angle of 1.6\,steradian,
a stellar magnetic field $B_*$=40~G as measured for the K-type star HD\,189733 (Moutou et al.\ 2007), 
and a stellar density at the base of the corona equal to the solar value, 
we obtain a predicted radio flux in this model of the order of 13\,mJy from 
HAT-P-11\,b, again similar to the tentative detection reported here. 

We conclude that a radio emission of the hot-Neptune HAT-P-11\,b at a level of 
a few milli-Jansky level is consistent with existing theoretical predictions, 
in particular the predictions based on the cyclotron maser instability (CMI)
of energetic electrons accelerated through the interaction of the planetary 
magnetic field with the stellar magnetic field, or coronal mass ejections.

\section{Conclusion}

Although we are not able to draw a definitive conclusion on  
150\,MHz radio emission from HAT-P-11\,b, at the very least the hint of radio detection 
presented here identifies HAT-P-11\,b as a prime candidate for many follow-up 
observations in the near future. The priority is to try to 
confirm the present tentative detection, 
via re-observation with GMRT at 150 MHz and/or new observations with LOFAR 
in the 30-250~MHz range (providing a 1-100~mJy sensitivity depending 
on the observation parameters used; van Haarlem et al., submitted, 2013) 
and UTR-2 in the 10-30~MHz range (providing a sensitivity of $\sim$100~mJy; 
Ryabov et al.\ 2004). 
The observed spectral range can also be extended toward shorter wavelengths using GMRT, 
with even higher sensitivities of $\sim$1~mJy at 240~MHz and $\sim$50~$\mu$Jy at 614~MHz 
(e.g., Lecavelier des Etangs et al.\ 2009). 
Observations over a broad range of frequencies will, with any luck, allow confirmation 
of the existence of the emission and better constraints on the planetary 
magnetic field strength and determination of the radio spectral index. 

The future observations should also be distributed at multiple epochs 
and at different orbital phases of the planet to 
characterize the suspected variability of the radio emission in terms of duty cycle, 
as well as the radio emission as a function of the star-planet angle and 
the emission directivity (CMI produces narrowly beamed radio emission). 
In this context, Hess \& Zarka (2011) have analyzed all the observables 
that could be derived 
from a broadband dynamic spectrum with a sufficient signal-to-noise ratio.

In the long term, it should be possible to investigate 
if the radio emission from extrasolar planetary systems 
correlates with the stellar spot activity. In this context, 
HAT-P-11 represents the target
of choice because the Kepler mission has recently demonstrated the possibility 
of following the spot activity and perhaps even
of drawing the butterfly diagram of surface spots for this star 
(Deming et al.\ 2011, Sanchis-Ojeda\& Winn 2011). 

Radio observations of different types of exoplanets are also crucial 
for comparative exoplanetology. For instance, observations of the other 
transiting hot-Neptune GJ436\,b (Gillon et al.\ 2007) 
will allow comparison with a planet similar in mass but orbiting 
an M-dwarf star which is half the mass and radius of HAT-P-11 and known to 
have a strong Far-UV and Lyman-$\alpha$ output (Ehrenreich et al.\ 2010). 
With a 6.55~Earth mass, 
the nearby exoplanet GJ1214\,b should also provide an opportunity to undertake 
search for 
radio emission from a water-rich lower mass super-Earth planet 
(Charbonneau et al.\ 2009; D\'esert et al.\ 2011). 
In the near future, more sensitive observations at even lower frequencies will 
become feasible with the LOFAR observatory, providing a great boost to 
exoplanetary research.  

\begin{acknowledgements}
This work is based on observations made with the Giant Meter Radio Telescope, 
which is operated by the National Centre for Radio Astrophysics of the 
Tata Institute of Fundamental Research and is located at Khodad, Maharashtra, India. 
We warmly thank the staff of the GMRT for the efficient support to these observations. 
We are especially indebted to Prof. Swarna K.\ Ghosh for allocating observation time from the
Director's discretionary quota. 
This work has been supported by the award of the Fondation Simone et Cino Del Duca. 
P.Z.'s activities in radio search for exoplanets are partly supported by 
the ANR program NT05-1-42530 ``Radio-Exopla''.

\end{acknowledgements}

\end{document}